\documentclass[fleqn,twoside]{article}
\usepackage{espcrc2}
\usepackage{epsfig}

\usepackage{graphicx}
\usepackage[figuresright]{rotating}

\newcommand{\AmS}{{\protect\the\textfont2A\kern-.1667em\lower.5ex\hbox{M}\kern-.125emS}}
\title{The polarised gluon density $\Delta G(x)$ from di-jet events at
high energy $ep$-colliders} 
\author{G. R\"adel\address[CEA]{DAPNIA/SPP, CEA Saclay, 91191 Gif-sur-Yvette,
France}\thanks{Present address: Laboratoire de M\'et\'eorologie Dynamique,
Ecole Polytechnique, 91128 Palaiseau, France} and
A. De Roeck\address{CERN, 1211 Geneva 23, Switzerland}
}

\begin{document}
\begin{abstract}
We present the potential to determine the polarised gluon density
from boson-gluon fusion processes with di-jet events
at future high energy $ep$-colliders.
These include HERA at DESY operated with polarised electrons
and protons, polarised protons from HERA colliding on polarised electrons 
from a future linear collider, and polarised 
protons from RHIC at BNL colliding on
electrons from a future electron accelerator.

\vspace{1pc}
\end{abstract}

\maketitle
\section*{Introduction}
The origin of the spin in the proton is still a subject of debate.
The last generation of deep inelastic scattering (DIS) experiments  has
 confirmed that the quarks
 account for only $30\%$ of the proton spin. 
Next-to-leading order (NLO) QCD fits of the $g_1$ structure function and 
semi-inclusive data suggest that the contribution of the gluon to the spin 
could be large~\cite{qcdfits}. A first attempt of a direct measurement of
 the polarised gluon distribution, $\Delta G$, using leading charged 
particles~\cite{hermes}
 is not in conflict with this suggestion. In general, it has
 been concluded that major progress in our understanding of the spin
 structure can be made with clear and unambiguous direct measurements of 
$\Delta G$.

\begin{figure}[htb]
\epsfxsize=6.5cm
\epsffile[50 60 380 200]{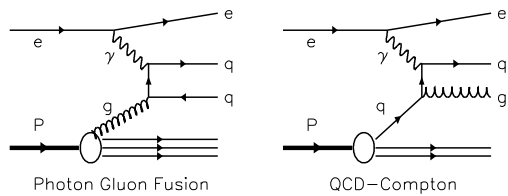}
\caption{Feynman diagrams for PGF and QCD-Compton processes.}
\label{fig:feynm}
\end{figure}

Polarized $ep$-colliders
 are particularly suited for this task. It has been demonstrated
by the present unpolarized studies at HERA that the large 
centre of mass system (CMS) energy allows
for several processes to be studied which  show a clear sensitivity to 
the gluon 
distribution in the proton. Consequently one could expect that
a powerful method to access $\Delta G$ is the measurement of 
di-jet rate asymmetries at high energy colliders of polarised electrons 
and protons.
In LO two diagrams can lead to di-jet events (Fig.~\ref{fig:feynm}):
 the
photon-gluon fusion 
(PGF) and the QCD-Compton process. The PGF is directly sensitive to the gluon 
density, while the QCD-Compton process is sensitive to the quark densities 
and constitutes the background.

Several feasibility studies on extracting the polarised gluon density 
$\Delta G$ 
from di-jet events at HERA in LO have been performed during the last 
years and are summarized in~\cite{gehr,works,gaby1}. These studies
established the di-jet process as a prime candidate to 
access the polarised gluon directly.
 Extracting $\Delta G$ at HERA in
NLO has been studied as well~\cite{heranlo}.
The studies have been extended to eRHIC and THERA. 
In this paper we summarize the main results of these studies.
In Section~1 we recall
the results for HERA.
Section~2 describes a first look at the potential of a combination of 
HERA and a possible
future linear collider at DESY. In Section~3 we show in a
somewhat greater detail what could be achieved at BNL, if a new
accelerator for polarised electrons would be  build, and its electrons
would collide with the polarised protons from RHIC.

\section{HERA as an $\vec{e}\vec{p}$-collider}
\begin{figure}[htb]
\epsfxsize=8.5cm
\epsffile[40 300 590 600]{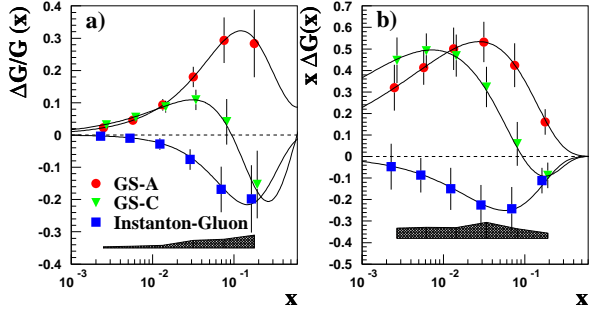}
\caption{Sensitivity to $\Delta G/G$ (a) and $x\Delta G$ (b) at HERA 
for a luminosity of $500~{\rm pb}^{-1}$ and three different assumptions
for the shape of $\Delta G(x)$. The error bars represent statistical
errors. The shaded band gives an estimate of the systematics. }
\label{fig:heralo}
\end{figure}

At HERA 820 (920) GeV protons collide  with 27.5 GeV electrons, providing
a center-of-mass energy  $\sqrt{s} \approx 300 (320)$~GeV.
Studies of upgrading HERA to a fully polarised $ep$-collider are discussed
in~\cite{barber,works}.
The potential of an extraction of $\Delta G$ in LO, assuming that both
beams
are  $70\%$ polarised, is shown in Figure~\ref{fig:heralo}.
The different assumptions for the particle density functions are
the Gehrmann-Stirling (GS) sets A and C~\cite{GS}, and the instanton-gluon
distribution~\cite{kochelev}. 

\begin{figure}[htb]
\epsfxsize=8.5cm
\epsffile[30 320 590 600]{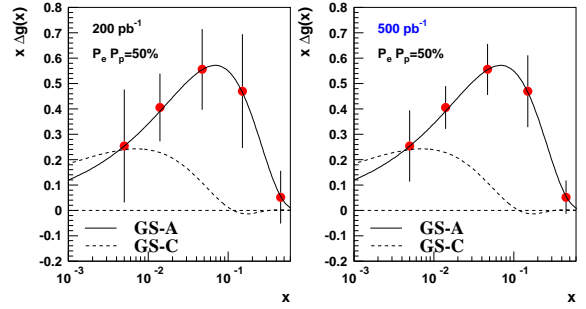}
\caption{The statistical precision of a measurement of $x\Delta G(x)$ from
di-jets at NLO, shown on top of the GS-A parton density curve, for two 
values of integrated luminosity. The expected value for the beam polarisation
is taken into account. GS-C is shown for reference.}
\label{fig:heranlo}
\end{figure}
For this study the Monte Carlo event generator PEPSI 6.5~\cite{PEPSI}
was used, which allowed to include hadronisation and detector effects.
Higher order effects were partly taken into account
 by unpolarised parton showers. The cuts chosen for this analysis are:
$5 < Q^2 < 100 {\rm GeV}^2$, $0.3 < y < 0.85$, 
and  two jets with a $p_t > 5$~GeV
are required in the pseudo-rapidity region of $|\eta_{jet}| < 2.8$.
Furthermore a cut on the invariant mass of the two jets, $s_{ij} > 100$~GeV,
is applied. Fig.~\ref{fig:heralo} demonstrates that the polarised 
gluon is 
 measurable in the $x$-range (of the gluon)  of  $0.002 < x < 0.2$. 
Note that this measurement allows the determination of the 
{\it shape} of $\Delta G(x)$. Furthermore it reaches $x$-values lower than
any other measurement planned in future so far, and (for a GS-A type of gluon)
will measure about 75\% of the first moment $\int \Delta G(x)~{\rm d}x$.
The errors on the individual points 
on $\Delta G(x)/G(x)$ are in the range of 0.007 to 0.1.
The total error on $\int \Delta G(x)/G(x)$ in the complete range is 0.02.

These studies have been extended to NLO~\cite{heranlo}, using the 
MEPJET~\cite{mepjet}
and DISENT~\cite{disent} programs. MEPJET is used since it includes
the polarised jet cross sections in NLO; DISENT is used to relate the 
$x_{jets}$ reconstructed from the measured jet pair to the true 
$x_g$ of the gluon. To improve the correlation between these
quantities, the cuts were increased: two jets with
 $p_t > 7$~GeV and  $s_{ij} > 200$~GeV are required. 
Generally the asymmetries are 
found to be  smaller than in the LO case and
 $x_g$-values get shifted towards larger $x$.
The results are shown in Fig~\ref{fig:heranlo}. The region of measurability
now becomes $0.005 < x < 0.5$, the measurement will be less precise than in LO,
but still has a very significant discrimination power, even already for
a luminosity of  200 pb$^{-1}$.
\begin{figure}[htb]
\epsfxsize=8.5cm
\epsffile[25 190 600 700]{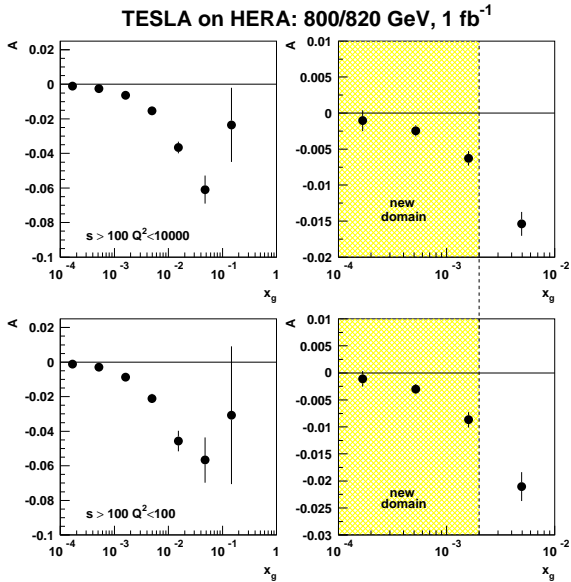}
\caption{Asymmetries measured using a 800 GeV $e^-$ beam of TESLA on
a 820 GeV $p$ beam from HERA, for two selected $Q^2$ regions (top/bottom).
On the left hand side the low-$x$ region is expanded and the newly reachable
 low-$x$ domain is shown by the hatched region of the plot.} 
\label{fig:thera}
\end{figure}
\section{THERA: combining HERA and TESLA}
THERA is a proposal to collide a 250 to 800~GeV electron beam, delivered by 
a linear collider such as TESLA, on the 920~GeV proton beam of 
HERA~\cite{thera}.
A di-jet study has been made in LO with MEPJET to check the reach of 
THERA~\cite{heranlo}.
The result is shown in Fig.~\ref{fig:thera} for jet selection
cuts as for (LO) HERA, and for the  optimistic case of
having  a 800~GeV electron beam on 820~GeV protons. The plot is shown
for a polarisation of the beams of 100\%.
 At the lowest $x$-values 
the asymmetries get very small, i.e.\ of order of 10$^{-3}$, 
and become unmeasurable even with a statistics of 1~fb$^{-1}$. But with a
statistics in that ball-park, one can hope to measure 
$\Delta G(x)$ about one order of magnitude lower in $x$ of the gluon
compared to HERA: 
the reach at THERA is  $0.0005 < x < 0.1$.

\section{eRHIC: adding electrons to RHIC}
Recently, it has been suggested to add an electron ring or electron LINAC
to the RHIC rings~\cite{erhic}. The latter  are presently being
commissioned to provide  a polarised proton beam of
250~GeV. The polarised 
electron beam energy would be around 10~GeV, leading to a 
total CMS energy of 100~GeV for this facility, 
baptised eRHIC, or EIC (Electron Ion Collider).
 The reachable luminosity is expected to be very high for this collider, 
of the order of 4~fb$^{-1}$/year.

A LO MEPJET study was made for eRHIC. The jet selection cuts have been 
optimized for reaching largest  sensitivity:
two jets with a $p_t > 3$~GeV
are required in the pseudo-rapidity region of $-3.5 < \eta_{jet} < 4$
and  $s_{ij} > 100$~GeV is required. 
The region in $\eta$ used is larger than for 
HERA and THERA, and will be a challenge for the detector design.
With HERA-like cuts some of the
precision is lost, but the measurement is still very significant.

\begin{figure}[htb]
\begin{center}
\epsfxsize=8cm
\epsffile[50 440 530 650]{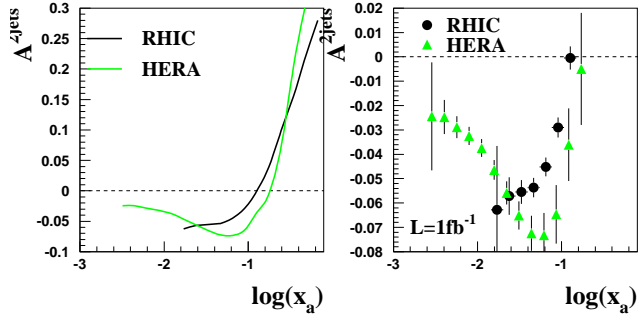}
\caption{Comparison of the di-jet asymmetry at eRHIC and HERA: 
(left) cross section curve, (right) measureable points for 
1 fb$^{-1}$.}
\label{fig:rhichera}
\end{center}
\end{figure}

\begin{figure}[htb]
\epsfxsize=8.5cm
\epsffile[25 300 580 580]{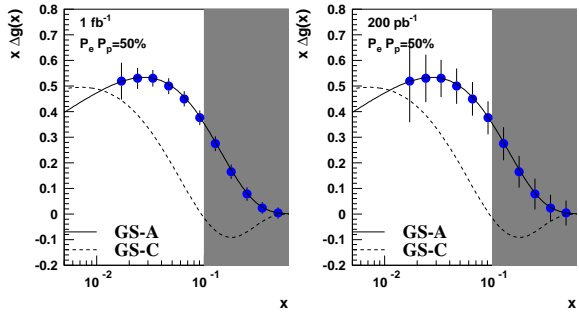}
\caption{The statistical precision of $x\Delta G$ from di-jets in LO
for eRHIC, for two different luminosities, with predictions of GS-A and
GS-C.}
\label{fig:glufig2}
\end{figure}

Fig.\ref{fig:rhichera} shows the asymmetries as function of 
$x_g$ for both HERA and eRHIC, and the expected error bars for 
1 fb$^{-1}$, calculated with 100\% polarisation. 
Fig.\ref{fig:glufig2} shows the statistical precision reachable 
for data samples with different luminosity, calculated with MEPJET,
and assuming $70\%$ polarisation for both beams.
The figure also 
shows the sensitivity to the gluon distribution. Clearly 
a good separation power between different models for $\Delta G$
is preserved.
The polarised gluon can be measured in the region
$0.02 < x < 0.3$. For larger values the asymmetries to the QCD compton 
background are dominating and its use will depend on how well one can control
this contribution.

In Table~\ref{tab2} we present the $x$-values of possible measurements
of $\Delta G$ and the corresponding sensitivities for two different
values of the $p_t$ cut of the jets. 
Hadronisation and detector effects are taken into account as correction
factors determined by comparing MEPJET and PEPSI results, as studied
 for HERA. These corrections inflate the errors by 40-60\%.
Lowering the jet $p_t$ is in 
particular important for reaching higher sensitivities in the low-$x$
region.

Finally we present a table with the sensitivities for HERA and for 
eRHIC of measurable points with statistical errors, after taking into 
account effects of NLO, hadronisation and detector smearing. 
All effects are  computed
as correction factors  to the LO MEPJET results,
determined from the detailed HERA studies with PEPSI and NLO
MEPJET/DISENT.
The result of the $x$-values of the possible data points and their error
is given in Table~\ref{tab1}. For eRHIC the values correspond to the
$p_t$-jets cut: $p_t>5$~GeV. These numbers can be used e.g. in 
general QCD fits to $g_1$ data, by using the di-jet cross sections
as a constraint on the gluon, as done e.g. in~\cite{abhay} for a 
LO study for a polarised 
HERA.
\begin{table}
\begin{center}
\caption{Possible measurements ($x$-values) and 
corresponding sensitivities to $\Delta
G$ at eRHIC in LO, for 4 fb$^{-1}$, for two different cuts on the $p_t$
of the jets. Beam polarisations of $P_e \cdot P_p =50\%$
are assumed.}

\begin{tabular}{|c|c||c|c|}
\multicolumn{4}{c}{} \\
\hline
\multicolumn{2}{|c||}{ $p_t >$ 3 GeV} & \multicolumn{2}{|c|}{ $p_t>5 $
GeV}
\\
\hline
       $ x$    &   $\delta( x \Delta G )$ & $x $    &  $\delta( x \Delta G )$\\
\hline
    0.017 &  0.057 & 0.017 &  0.332 \\
    0.024 &  0.032 & 0.024 &  0.208 \\
    0.033 &  0.025 & 0.033 &  0.063 \\
    0.047 & 0.023  & 0.047 &  0.045\\
    0.065 & 0.022  & 0.065 &  0.036\\
    0.090 & 0.020  & 0.090  & 0.030 \\
    0.13 &  0.020    & 0.13  &  0.028 \\
    0.18 &  0.019  & 0.18 &   0.026\\
    0.25 &  0.019  & 0.25  &  0.025 \\
\hline
\hline
\end{tabular}
\label{tab2}
\end{center}
\end{table}
\begin{table}
\begin{center}
\caption{Possible measurements ($x$-values) and 
corresponding sensitivities to $\Delta
G$ at eRHIC and HERA in NLO. ($P_e \cdot P_p =50\%$).}
\label{tab1}
\begin{tabular}{|c|c||c|c|}
\multicolumn{4}{c}{} \\
\hline
\multicolumn{2}{|c||}{ HERA (500 pb$^{-1}$)} & \multicolumn{2}{|c|}{ eRHIC (4 fb$^{-1}$)} \\
\hline
       $ x$    &   $\delta( x \Delta G )$ & $x $    &  $\delta( x \Delta G )$\\
\hline
    0.005 &  0.200 & 0.024 &  0.180 \\
    0.014 &  0.112 & 0.033 &  0.110 \\
    0.047 &  0.140 & 0.047 &  0.084 \\
    0.15  &  0.200 &0.065 &  0.068\\
    0.45 &  0.092 & 0.090 &  0.060\\
& & 0.13  &  0.058 \\
& & 0.18  &  0.053 \\
& & 0.25  &  0.049 \\
\hline
\hline
\end{tabular}

\end{center}
\end{table}

\section{Summary}

\begin{figure}[htb]
\epsfxsize=7.5cm 
\epsffile[40 200 550 680]{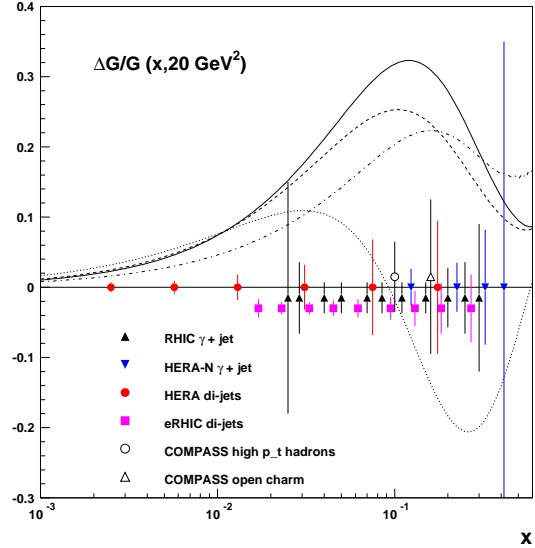}
\caption{Summary of the sensitivity of future measurements
of $\Delta G/G$ for HERA (500 fb$^{-1}$) and eRHIC (4 fb$^{-1}$)
compared with other  experiments (see text).}
\label{fig:sumup}
\end{figure}

In summary
di-jet measurements at a future polarised $ep$-collider will give
important information on the shape and magnitude of the polarised
gluon  distribution of the proton. For all three $ep$-collider
versions studied, these measurements are found to be feasible. 
The global result is shown in Fig.~\ref{fig:sumup}, for HERA and 
eRHIC and compared to results from other planned or possible 
future polarised experiments: $pp$-scattering at RHIC~\cite{saito}
(STAR, $\sqrt{s}=200$ GeV, ${\cal L} = 320~{\rm pb}^{-1}$),
$\mu p$-scattering in COMPASS~\cite{kunne}, and $pp$-scattering in
HERA-N~\cite{nowak}, i.e. using the polarised protons of HERA on a 
polarised fixed target. All measurements are shown at LO. 
Clearly eRHIC can produce the (statistically) most precise 
measurements, and HERA covers the range to lowest possible $x$-values 
which can only be beaten by THERA, if it would be able to deliver 
sufficient luminosity.

In all di-jet measurements at $ep$-colliders appear to be a decisive tool
to analyse and settle the question on $\Delta G$.

\end{document}